\PassOptionsToPackage{english}{babel}
\PassOptionsToPackage{svgnames}{xcolor}
\documentclass[sigplan,screen]{acmart}
\makeatletter\let\@authorsaddresses\@empty\makeatother
\usepackage[utf8]{inputenc}
\usepackage[T1]{fontenc}
\usepackage[noframes]{ffcode}
\usepackage{anyfontsize}
\usepackage{booktabs}
\usepackage{caption}
\usepackage{csquotes}
\usepackage{doi}
\usepackage{amsmath}
\usepackage{eolang}
\usepackage{hyperref}
\usepackage{mathtools}
\usepackage{mdframed}
\usepackage{multicol}
\usepackage{naive-ebnf}
\usepackage{nameref}
\usepackage{paralist}
\usepackage{semantic}
\usepackage{subcaption}
\usepackage{tabularx}
\usepackage{tcolorbox}
\usepackage{tikz}
  \usetikzlibrary{backgrounds}
  \usetikzlibrary{fit}
  \usetikzlibrary{arrows.meta}
\usepackage{textcomp}
\usepackage{to-be-determined}
\usepackage{url}
\usepackage{xurl}
\usepackage[shortlabels]{enumitem} 
\usepackage[capitalize,nameinlink]{cleveref} 

\copyrightyear{2026}
\acmYear{2026}
\setcopyright{cc}
\setcctype{by}
\acmConference[SOAP '26]{Proceedings of the 15th ACM SIGPLAN International Workshop on the State Of the Art in Program Analysis}{June 15--19, 2026}{Boulder, CO, USA}
\acmBooktitle{Proceedings of the 15th ACM SIGPLAN International Workshop on the State Of the Art in Program Analysis (SOAP '26), June 15--19, 2026, Boulder, CO, USA}
\acmDOI{10.1145/3814987.3814988}
\acmISBN{979-8-4007-2709-2/2026/06}
\acmSubmissionID{pldiws26soapmain-p1-p}
\received{2026-02-10}
\received[accepted]{2026-04-15}

\title{Compile-Time Java Stream Fusion via \texttt{mapMulti}}

\author{Yegor Bugayenko}
\orcid{0000-0001-6370-0678}
\email{yegor256@huawei.com}
\affiliation{\institution{Huawei}\city{Shenzhen}\country{China}}
\author{Maxim Trunnikov}
\orcid{0000-0001-8545-4797}
\email{trunnikov.maxim@huawei.com}
\affiliation{\institution{Huawei}\city{Shenzhen}\country{China}}
\author{Vladimir Zakharov}
\orcid{0009-0007-2420-5675}
\email{vladimir.zakharov@huawei.com}
\affiliation{\institution{Huawei}\city{Shenzhen}\country{China}}


\newcommand\figcap[1]{%
  \Description{#1}%
  \caption{#1}%
}
\newcommand\pragma[1]{\texttt{\texorpdfstring{\(\Phi\)}{Q}.#1}}
\newcommand\novale[2]{#2}
\newcommand\highlight[1]{{\color{orange}\ifmmode\text{#1}\else#1\fi}}
\newcommand\hone{\anon[T1]{{\scshape Hone}}}

\newcommand\streamliner{{\scshape Streamliner}}
\newcommand\jeo{\anon[T2]{{\scshape Jeo}}}
\newcommand\phino{\anon[T3]{{\scshape Phino}}}
\newcommand\requ[1]{\textbf{RQ#1}}

\makeatletter
\newcommand\frg[2]{%
  \def\param{#1}%
  {\overbracket[0.4pt][1pt]{#2}^{{\ifx\param\empty\else\color{gray}\text{\sffamily#1}\fi}}}}
\makeatother

\newlength{\firstsf}
\newlength{\secondsf}
\newlength{\thirdsf}
\newcommand\settwocapwidths[1]{
  \setlength{\firstsf}{#1\linewidth}
  \setlength{\secondsf}{\linewidth - \firstsf - 1em}}
\newcommand\setthreecapwidths[2]{
  \setlength{\firstsf}{#1\linewidth}
  \setlength{\secondsf}{#2\linewidth}
  \setlength{\thirdsf}{\linewidth - \firstsf - \secondsf - 1em}}
\newcommand\twofigcaps[4]{
  \begin{subfigure}[t]{\firstsf}
  \figcap{#2}
  \label{fig:#1-left}
  \end{subfigure}
  \hfill
  \begin{subfigure}[t]{\secondsf}
  \figcap{#3}
  \label{fig:#1-right}
  \end{subfigure}
  \figcap{#4}
  \label{fig:#1}}
\newcommand\threefigcaps[5]{
  \begin{subfigure}[t]{\firstsf}
  \figcap{#2}
  \label{fig:#1-left}
  \end{subfigure}
  \hfill
  \begin{subfigure}[t]{\secondsf}
  \figcap{#3}
  \label{fig:#1-middle}
  \end{subfigure}
  \hfill
  \begin{subfigure}[t]{\thirdsf}
  \figcap{#4}
  \label{fig:#1-right}
  \end{subfigure}
  \figcap{#5}
  \label{fig:#1}}

\begin{abstract}

The Java Stream API, introduced in Java~8, makes data processing more expressive and concise compared to imperative loops.
However, this abstraction can come with significant performance overhead, often due to the creation of multiple intermediate objects during pipeline execution.
In functional languages such as Haskell, this problem is addressed through stream fusion, a compile-time optimization that eliminates unnecessary intermediate structures.
Inspired by this idea, \streamliner{} was the first tool to perform ahead-of-time, bytecode-to-bytecode stream optimization for Java by unrolling stream pipelines into imperative loops.
In this paper, we introduce an open-source optimizer that takes a different approach.
Instead of unrolling streams into loops, it merges consecutive \ff{map()} and \ff{filter()} operations into a single \ff{mapMulti()} call, available since Java 16.
Our method avoids several limitations of \streamliner{}, including its sensitivity to escaping objects in lambda expressions and its restrictions on assigning or passing streams as variables.
We evaluated our optimizer on nine benchmarks and observed superior performance in two cases and comparable results in most others.
We also applied it to the bytecode of Apache Kafka, successfully executing all 31,799 unit tests without failures.

\end{abstract}

\begin{document}

\raggedbottom

\maketitle

\newcommand\sctn[2]{
  \section{#1}
  \label{sec:#2}
  \input{tex/#2}}

  \section{Introduction}
  \label{sec:introduction}

The Stream API\footnote{\url{https://openjdk.org/projects/jdk8/}} introduced in Java~8 makes stream processing more expressive~\citep{kiselyov2017}.
However, it relies heavily on virtual calls to temporary objects.
Consequently, streams may work slower than procedural imperative loops~\citep{khatchadourian2020study}.

Stream fusion may help by reducing stream operations and virtual calls~\citep{hirzel2014catalog}.
Consider the following Java stream pipeline:
\begin{ffcode}
var NN = new int[] {1, 1, 2, 3, 5, 8, 13};
var s = IntStream.of(NN) (*@\label{ln:stream-start}@*)
  .map(x -> (*@\highlight{x * x}@*))
  .map(x -> (*@\highlight{x + 1}@*))
  .sum(); (*@\label{ln:stream-end}@*)
\end{ffcode}
It can be optimized by joining two \ff{map} operations into one:
\begin{ffcode}
var s = IntStream.of(NN).map(
  x -> (*@\highlight{x * x + 1}@*)).sum();
\end{ffcode}
It can be further unrolled into an imperative loop:
\begin{ffcode}
var s = 0;
for (var i = 0; i < NN.length; i++)
  s += NN[i] * NN[i] + 1;
\end{ffcode}

The first ahead-of-time (AOT) bytecode-to-bytecode optimizer called \streamliner{} was introduced by \citet{moller2020eliminating}.
It \emph{unrolls} some stream pipelines by transforming them into loops, thus eliminating virtual calls.
However, this approach has limitations.
For example, objects may not escape from lambda expressions.
Additionally, streams may not be assigned to variables or passed as method arguments.
Moreover, \streamliner{} increased performance in older JVMs, but we couldn't reproduce similar results in JVMs newer than 17.
Our experiments showed that \streamliner{} may \emph{reduce} performance of short pipelines in recent JVM versions.

In Java~16, the Stream API was extended with the \ff{mapMulti} method, which works similarly to \ff{flatMap} but is an order of magnitude faster.
We implemented stream \emph{fusion} by combining pairs of \ff{map} and \ff{filter} operations into a single \ff{mapMulti} operation.
We don't replace streams with loops but merge multiple stream operations into a single one.
The snippet at the lines \ref{ln:stream-start}--\ref{ln:stream-end} would look like this, after our optimization:

\begin{ffcode}
var s = IntStream.of(NN).mapMulti(
  (x, c) -> { (*@\highlight{c.accept(x * x + 1)}@*); }
).sum();
\end{ffcode}

Instead of direct bytecode manipulation, our optimization pipeline is implemented as a rewriting system.
We disassemble bytecode to \phic{} expressions~\citep{bugayenko2021eolang}, rewrite them, and assemble into bytecode.
The usage of \phic{} enabled bytecode manipulation through declarative pattern matching and rewriting.

We released \hone{} to Maven Central, an open source Maven plugin that implements our method%
  \footnote{\anon[URL anonymized]{\url{https://github.com/objectionary/hone-maven-plugin}}}.
Any Java project may optimize its bytecode after compilation by adding our plugin to their Maven or Gradle build configuration.
The plugin requires no modifications to source code.

We asked the following research questions:
\requ{1}: Can replacing sequences of \ff{map} and \ff{filter} operations with a single \ff{mapMulti} operation provide performance gains similar to stream unrolling?
\requ{2}: Can stream fusion via \ff{mapMulti} be applied safely, not disrupting the functionality of existing stream pipelines?

Our results were as follows:
On two of nine benchmarks used by \citet{moller2020eliminating}, our optimization outperformed \streamliner{}.
On five benchmarks, our optimization matched \streamliner{}.
Our method didn't optimize two benchmarks because they use terminal stream operations that our method is not intended to optimize.
Optimization of Apache Kafka bytecode didn't cause any of its 31,799 unit tests in 1,511 test suites to fail.

Based on the results, we can provide the following answers to our RQs:
First, stream fusion via \ff{mapMulti} can provide performance gains comparable to or better than stream unrolling.
Second, our optimization has fewer applicability limitations in production Java systems compared with \streamliner{}.

In summary, the contributions of this paper are:
\begin{inparaenum}[1)]
  \item We propose a formal intermediate representation for Java bytecode based on \phic{}, enabling systematic stream fusion via rewrite rules and bridging low-level bytecode with high-level algebraic optimization;
  \item We present \jeo{} v0.15.1 , a Maven plugin for bidirectional mapping between Java bytecode and \phic{} expressions%
      \footnote{\anon[URL anonymized]{\url{https://github.com/objectionary/jeo-maven-plugin}}};
    a rewriting system with 44 rules that replaces \ff{map} and \ff{filter} operations with \ff{mapMulti} operations;
    \phino{} v0.0.0.67 , a tool that applies rewriting rules to \phic{} expressions%
      \footnote{\anon[URL anonymized]{\url{https://github.com/objectionary/phino}}};
    and \hone{} v0.24.0 , a Maven plugin that automatically optimizes bytecode \ff{.class} files after compilation;
  \item We report an experimental evaluation showing that our method of stream fusion
    may be more effective on some benchmarks than \streamliner{}, especially on modern JVMs.
  \item We demonstrate by experiment that our method doesn't disrupt
    the functionality of 31,799 unit tests in Apache Kafka.
\end{inparaenum}

All our digital artifacts, including the results of experiments and all used software, are published in Zenodo~\cite{zenodo2026}.

  \section{Background}
  \label{sec:background}

Java 8 introduced lambda expressions and streams in 2014.
Streams make code more expressive compared to traditional imperative loops~\citep{khatchadourian2020study}.
Although they resemble functions from functional programming, they are objects, not functions.
The Stream API consists of regular Java classes, while the lambda expressions passed to stream operations are compiled to \ff{invokedynamic} instructions and corresponding static methods.
At runtime, when the JVM encounters an \ff{invokedynamic} instruction, it creates an instance of a class that implements the target functional interface.
All calls to the lambda expression then become virtual calls to that created instance.

Streams in Java can be slow due to their abstraction overhead~\citep{kiselyov2017,khatchadourian2020study}.
Each \ff{map} and \ff{filter} operation creates a new stream with its own infrastructure.
For a simple sum of five squares, the stream pipeline invokes the lambda function five times, plus additional overhead from stream machinery:
\begin{ffcode}
var NN = {1, 2, 3, 4, 5};
var s = IntStream.of(NN).map((*@\highlight{x -> x * x}@*)).sum();
\end{ffcode}
This algorithm can be rewritten as an imperative loop that doesn't use virtual calls:
\begin{ffcode}
var s = 0;
for (var i = 0; i < NN.length; i++)
  s += NN[i] * NN[i];
\end{ffcode}

Performance advantages of the imperative solution were obvious in earlier JVMs.
However, modern JVMs are different.

\begin{figure*}
\settwocapwidths{0.45}
\begin{subfigure}[t]{\firstsf}
\begin{ffcode}
public long loop() {
  long sum = 0L;
  (*@\highlight{for}@*) (int i=0; i<NN.length; ++i)
    if (NN[i] 
      sum += NN[i] * NN[i];
  return sum; }
\end{ffcode}
\end{subfigure}
\hfill
\begin{subfigure}[t]{\secondsf}
\begin{ffcode}
public long map() {
  return IntStream.of(NN)
    .(*@\highlight{filter}@*)(x -> x 
    .(*@\highlight{map}@*)(x -> x * x)
    .(*@\highlight{sum}@*)(); }
\end{ffcode}
\end{subfigure}
\twofigcaps
  {moller}
  {Imperative style.}
  {Functional style, using a stream pipeline.}
  {Two variants of computing sums of even squares suggested by \citet{moller2020eliminating} to demonstrate that the usage of the Stream API degrades performance.}
\end{figure*}

\begin{table}
\setlength{\tabcolsep}{.3em}
\figcap{%
  The ``Ratio'' column shows stream time divided by loop time.
  It compares \cref{fig:moller-left} with \cref{fig:moller-right} for each JVM version.
  The numbers demonstrate that the performance of streams was improved since earlier JVM versions.}
  \begin{center}%
  \begin{tabularx}{\linewidth}{X|l>{\ttfamily}r>{\ttfamily}r|l>{\ttfamily}r>{\ttfamily\arraybackslash}r}
\toprule
Vendor & JVM & {\rmfamily Ratio} & {\rmfamily RSD} & JVM & {\rmfamily Ratio} & {\rmfamily RSD} \\
\midrule
    Corretto & 8.0.442 & \textcolor{DarkRed}{1.86} & 1.8\% & 24.0.2 & 0.83 & 1.0\% \\ 
    GraalVM & 17.0.12 & \textcolor{DarkGreen}{0.61} & 1.1\% & 26.ea.6 & \textcolor{DarkGreen}{0.56} & 0.9\% \\ 
    Java.net & 21.0.2 & 0.82 & 2.2\% & 26.ea.7 & 0.83 & 7.5\% \\ 
    Microsoft & 11.0.26 & 0.96 & 0.9\% & 21.0.8 & 0.82 & 1.9\% \\ 
    Oracle & 17.0.12 & 0.82 & 2.4\% & 24.0.2 & 0.83 & 1.3\% \\ 
    Temurin & 11.0.18 & 0.98 & 5.2\% & 24.0.2 & 0.84 & 1.4\% \\ 
    Tencent & 8.0.432 & \textcolor{DarkRed}{1.86} & 2.7\% & 21.0.7 & 0.83 & 1.5\% \\ 
    Zulu & 8.0.462 & \textcolor{DarkRed}{1.87} & 1.0\% & 24.0.2 & 0.83 & 1.2\% \\ 
\bottomrule
\end{tabularx}
  \label{tab:moller-jvms}%
  \end{center}
\end{table}

Modern JVMs handle streams better than their earlier versions.
We reproduced the experiment of \citet{moller2020eliminating}.
\Cref{fig:moller} shows two code snippets they tested with Java~13 using the OpenJDK Server~VM (build 13+33).
They observed a ``60\% performance degradation'' for the stream approach (\cref{fig:moller-right}) compared to the loop (\cref{fig:moller-left}).
\cref{tab:moller-jvms} shows the data we collected: the degradation is present only in versions older than Java~11.
Moreover, some of the latest JVMs calculate the sum of even numbers faster when using a stream of integers.

In this and all other experiments, we used JMH\footnote{\url{https://github.com/openjdk/jmh}} with
  ten  warming-up and ten  measuring iterations.
We used
  a bare metal
  macOS 15.7.4 

  at
  3.5 
~GHz
  arm64 

  machine
  with 12 
~cores
  and 32 
~GiB of memory.
The maximum measurement error observed in all experiments was 0.059 \%.
The size of the \ff{NN} array was 100  million.
The relative standard deviation (RSD), here and in all other tables in the paper, is a measure of relative variability.
We performed all experiments ten  times and calculated RSD from the standard deviation and mean.
In all our measurement experiments we used JVMs installable via SdkMan\footnote{\url{https://sdkman.io/}}.

Under the hood, modern HotSpot JVMs have dramatically improved their JIT compilers and runtime machinery compared to the Java~8 era.

First, their inlining is more aggressive.
In Java~8, the C2 compiler's default MaxInlineLevel was 9, limiting nested call inlining.
By JDK~11, the inlining budget and heuristics were refined.
They better handle functional patterns.
Small lambdas and short stream pipeline operations are inlined and optimized into efficient loops.

Second, they improved escape analysis and scalar replacement.
Intermediate objects created by the Stream API can now be optimized away.
If the JIT proves that they never escape the method boundary, it eliminates the allocation entirely through scalar replacement.
This transforms what used to be dozens of fleeting allocations and GC events into pure register operations.

By contrast, a manually written loop often presents less uniform control flow.
It hides opportunities for scalar replacement and may not be profiled as comprehensively by the compiler.
Short pipelines---once vilified for their ``object churn''---now run as fast as or faster than equivalent for-loops.

  \section{Optimization Pipeline}
  \label{sec:pipeline}

To reason about and transform bytecode structurally, we encode it as \(\varphi\)-expressions,
  a minimal, uniform representation that exposes both instructions and metadata in a single composable form.
The full calculus was introduced by \citet{bugayenko2021eolang}; we use a subset sufficient to represent Java bytecode.

\Cref{fig:jeo} shows a stream pipeline compiled to bytecode and then disassembled into a \(\varphi\)-expression.
Because \phic{} uses composable symbolic structures, its expressions are amenable to rule-based rewriting.
For example, the following rule transforms \ff{IMUL} to \ff{IADD}, where \(B_1\) and \(B_2\) match any sequence of key–value pairs:
\begin{phiquation*}
 \frg{pattern}{ [[ B_1, @ -> Q.imul, B_2 ]] } \longrightarrow \frg{replacement}{ [[ B_1, @ -> Q.iadd, B_2 ]] }.
\end{phiquation*}
Rules may use functions from \phino{}, our rewriting engine.
The following rule renames classes matching a pattern:
\begin{phiquation*}
 [[ B_1, name -> e_1, B_2 ]] \longrightarrow [[ B_1, name -> e_2, B_2 ]]
   \qquad\qquad\text{when}\quad |matches| {(} |"\char94Nums\textdollar{}"| {,} e_1 {)}
   \qquad\qquad\text{where}\quad e_2 := |concat| {(} |"Op"| {,} |name| {)} {.}
\end{phiquation*}
Rules are applied repeatedly until none match; order is irrelevant.

We scan bytecode files using the ASM library\footnote{\url{https://asm.ow2.io/}} and disassemble them into \(\varphi\)-expressions (one \ff{.class} file produces one \ff{.phi} file).

\begin{figure*}
\setthreecapwidths{0.31}{0.27}
\begin{subfigure}[t]{\firstsf}
\begin{ffcode}
import
  java.util.stream.*;
class Numbers {
 int total(int[] a) {
  return IntStream
   .of(a)
   .map((*@\highlight{x -> x * x}@*))
   .sum();
 }
}
\end{ffcode}
\end{subfigure}
\hfill
\begin{subfigure}[t]{\secondsf}
\begin{ffcode}
public Numbers():
 aload_0
 invokespecial
 return
public int
  total(int[]):
 aload_1
 invokestatic
 invokedynamic
 invokeinterface
 invokeinterface
 ireturn
private static int
  (*@\highlight{total\textdollar{}0}@*)(int):
 iload_0
 iload_0
 imul
 ireturn
\end{ffcode}
\end{subfigure}
\hfill
\begin{subfigure}[t]{\thirdsf}
\begin{phiquation*}
[[ |j\textdollar{}Numbers| -> [[
  @ -> Q.class,
  version -> 61,
  access -> 33,
  supername ->
    "java/lang/Object",
  interfaces -> [[ ]],
  |j\textdollar{}object\char64init\char64| -> [[\dots]],
  |j\textdollar{}total| -> [[\dots]],
  |\highlight{j\textdollar{}total\textdollar{}0}| -> [[
    @ -> Q.method,
    access -> 4106,
    descriptor -> "(I)I",
    body -> [[
      @ -> Q.seq.of6,
      i2 -> [[ @ -> Q.iload, |v| -> 0 ]],
      i3 -> [[ @ -> Q.iload, |v| -> 0 ]],
      i4 -> [[ @ -> Q.imul ]],
      i5 -> [[ @ -> Q.ireturn ]] ]] ]] ]] ]]
\end{phiquation*}
\end{subfigure}
\threefigcaps
  {jeo}
  {Java class.}
  {Bytecode, from \ff{javap}.}
  {Simplified \(\varphi\)-expression.}
  {Java class after compilation and disassembly to \(\varphi\)-expression.}
\end{figure*}

Each instruction becomes a formation qualified by a dispatch (e.g., $Q.class$ for classes, $Q.method$ for methods); parameterized instructions like \ff{ILOAD} carry their parameters as attributes.
Java names are prefixed with \ff{j\textdollar{}} to distinguish them from meta-attributes.

The Java compiler translates lambda expressions into \ff{invokedynamic} and \ff{invokeinterface} instruction pairs~\citep{rose2009bytecodes}.
We re\-write these pairs into \pragma{filter} or \pragma{map} pragmas---\phic{} formations that temporarily carry rewriting data but don't exist in the final bytecode.

Both \pragma{filter} and \pragma{map} are then unified into \pragma{distill}, an abstraction accepting an item and returning \ff{void}.
Consecutive \pragma{distill} pragmas are merged by concatenating their lambda bodies.
Finally, \pragma{distill} pragmas become \pragma{mapMulti} pragmas, which generate new private static methods containing the merged logic plus a call to the \ff{Consumer}'s \ff{accept} method.
The pragmas are then transformed back into instructions and reassembled into bytecode via ASM.

  \section{Results}
  \label{sec:results}

\textbf{RQ1: \texttt{mapMulti()} worked faster than \texttt{map()} + \texttt{filter()}}.
We used the same benchmark as \citet{moller2020eliminating}, since other Java benchmarks (SPECjbb~2015, DaCapo, Renaissance) either don't use the Stream API or contain only trivial pipelines.
\Cref{tab:streamliner-vs-hone} shows JMH measurements on two Zulu JVM versions; ratios below \(1.0\) indicate speedup.

\begin{table*}
\figcap{Optimization effect of \streamliner{} (``{\scshape Strm}'' column) compared with our method (``\hone{}'' column) on two Zulu JVM versions. Each ratio divides optimized time by original time.}
  \begin{center}%
  \begin{tabularx}{\linewidth}{X|>{\ttfamily}r>{\ttfamily}r>{\ttfamily}r>{\ttfamily}r|>{\ttfamily}r>{\ttfamily}r>{\ttfamily}r>{\ttfamily\arraybackslash}r}
\toprule
Benchmark & \multicolumn{4}{c|}{Zulu 17.0.16}  & \multicolumn{4}{c}{Zulu 23.0.2} \\
  & {\scshape Strm} & \(\pm\)RSD & \hone{} & \(\pm\)RSD  & {\scshape Strm} & \(\pm\)RSD & \hone{} & \(\pm\)RSD\\
\midrule
  \texttt{allMatch} & \textcolor{DarkGreen}{0.12} & 13.9\% & 1.00 & 0.0\% & \textcolor{DarkGreen}{0.17} & 0.0\% & 1.00 & 0.0\%\\
  \texttt{count} & \textcolor{DarkGreen}{0.11} & 4.9\% & 1.00 & 8.2\% & \textcolor{DarkGreen}{0.20} & 0.0\% & 1.00 & 0.0\%\\
  \texttt{filterCount} & \textcolor{DarkRed}{2.07} & 0.4\% & 1.01 & 2.8\% & \textcolor{DarkRed}{2.04} & 2.1\% & 0.99 & 2.1\%\\
  \texttt{filterMapCount} & \textcolor{DarkRed}{2.13} & 2.7\% & 1.02 & 2.9\% & \textcolor{DarkRed}{2.12} & 1.6\% & 1.02 & 1.7\%\\
  \texttt{megamorphicFilters} & \textcolor{DarkGreen}{0.05} & 0.7\% & \textcolor{DarkGreen}{0.05} & 0.5\% & \textcolor{DarkGreen}{0.05} & 2.1\% & \textcolor{DarkGreen}{0.05} & 0.8\%\\
  \texttt{megamorphicMaps} & \textcolor{DarkGreen}{0.01} & 0.5\% & \textcolor{DarkGreen}{0.03} & 0.4\% & \textcolor{DarkGreen}{0.01} & 1.5\% & \textcolor{DarkGreen}{0.01} & 1.8\%\\
  \texttt{sum} & 1.00 & 0.8\% & 1.00 & 0.7\% & 0.99 & 0.6\% & 1.00 & 1.0\%\\
  \texttt{sumOfSquares} & \textcolor{DarkGreen}{0.29} & 0.6\% & 1.00 & 0.5\% & 1.00 & 1.0\% & 1.00 & 0.6\%\\
  \texttt{sumOfSquaresEven} & 1.02 & 0.5\% & 1.01 & 0.3\% & 0.99 & 3.4\% & 0.98 & 3.3\%\\
\bottomrule
\end{tabularx}
  \label{tab:streamliner-vs-hone}%
  \end{center}
\end{table*}

On the latest JVM, three benchmarks (\ff{sum}, \ff{sumOfSquares}, \ff{sumOfSquaresEven}) showed no improvement from either method.
Two benchmarks (\ff{megamorphic*}) ran faster with both.
Two benchmarks (\ff{allMatch}, \ff{count}) improved only with \streamliner{}, as our method doesn't modify these terminal operations.
Notably, two benchmarks (\ff{filterCount}, \ff{filterMapCount}) ran slower after \streamliner{} but showed no degradation with our method.
Overall, our optimization either preserved speed or produced up to 5x improvements, with no performance degradation.

\textbf{RQ2: The optimization didn't disrupt existing functionality}.
We applied \hone{} to Apache Kafka (tag~4.0.0): 992k lines of Java, 142k lines of Scala, 31,799 unit tests.
After disabling 13 flaky tests, the original build took 120 minutes; with \hone{} injected into \ff{build.gradle}, it took 165 minutes and affected 15k \ff{.class} files.
All tests passed without additional modifications.
We also integrated \hone{} into its own Maven release pipeline; after 50+ releases, no functionality was disrupted.
The modified Kafka is available at: \anon[URL anonymized]{\url{https://github.com/objectionary/kafka}}.

  \section{Related Work}
  \label{sec:related}

Eliminating intermediate structures in functional programs, known as ``deforestation'' or fusion~\citep{hirzel2014catalog}, has been studied since the late \novale{Google.Units}{1980s}~\citep{wadler1990deforestation} and implemented in compilers such as GHC~\citep{gill1993fusion}.
A common approach is deductive fusion~\citep{bird1989calculation,meijer1991bananas,chin1992fusion,takano1995deforestation,bird1997algebra,hamilton2002deforestation,coutts2007,hinze2011fusion}, which applies rewrite rules to produce optimized programs.
Similar techniques have been applied to OCaml and Scala through staging~\citep{kiselyov2017}, and recent work on \textsc{SuperFusion}~\citep{ji2024superfusion} relies on supercompilation.

In object-oriented languages, \citet{kiselyov2017} proposed stream fusion via \textit{strymonas} for Scala and OCaml, though it cannot be adapted to Java due to missing metaprogramming capabilities~\citep{kiselyov2024complete}.
\citet{ribeiro2018java} introduced a Java library transforming streams into loops, but both approaches require switching from the standard Stream API.
\citet{khatchadourian2020safe} developed an Eclipse plug-in for detecting parallelizable streams, observing 3.5x speedups, while \citet{basso2022optimizing} proposed stream parallelization for optimization.
\citet{moller2020eliminating} introduced compile-time stream unrolling via \streamliner{}, demonstrating improvements on benchmarks by \citet{biboudis2014clash}.
Similar solutions exist for C\# (Steno~\citep{murray2011steno}) and F\# (LinqOptimizer).

For bytecode manipulation, frameworks such as ASM, Byte Buddy, Javassist, and Apache Commons BCEL provide imperative APIs requiring manual traversal.
Analysis-orien\-ted tools translate bytecode into intermediate representations: Doop uses Datalog facts, WALA~IR~\citep{santos2022program} provides SSA-style analysis, OPAL~\citep{eichberg2014software} and TACAI~\citep{reif2020tacai} offer three-address code, and Tai-e~\citep{tian23tai} focuses on static analysis.
Soot~\citep{vallee1999soot} provides multiple IRs (Baf, Jimple, Shimple, Grimp) for transformation, while GraalVM~\citep{duboscq2013graal} uses internal graph-based IRs for compilation.
Krakatau offers textual bytecode encoding without formal semantics.
All these tools rely on imperative transformation or analysis-orien\-ted IRs without declarative rewriting support.
In contrast, \phino{} enables developers to specify rewriting rules that the system applies automatically, shifting effort from imperative traversal to concise, formally specified rewrites.

  \section{Limitations}
  \label{sec:limitations}

Our method requires JDK~16 or newer due to \ff{mapMulti}, is hardwired to the Stream API tested on JDK~24, and currently transforms only \ff{map} and \ff{filter} among the eleven intermediate operations (\ff{dropWhile}, \ff{flatMap}, \ff{gather}, \ff{limit}, \ff{peek}, \ff{skip}, \ff{sorted}, \ff{takeWhile}, etc.).
We cannot fuse closures --- lambdas capturing outer variables --- because bytecode places \ff{ILOAD} and \ff{ALOAD} instructions between stream operations, breaking the consecutive pattern our rewriting rules expect.
Similarly, we cannot fuse functional objects (e.g., \ff{new Square()} instead of lambdas) as they compile to \ff{new}/\ff{invokespecial} rather than \ff{invokedynamic}.
Code relying on Reflection to inspect bytecode structure may break, though such practice is uncommon.
Line metadata may be damaged, affecting debugging.

Build time increased by 45 minutes (37\%) for Apache Kafka; merging the current two-component architecture (Java+XSLT and Haskell) into a unified stack may improve this.
We lack micro-level profiling (allocation counts, JIT inlining traces) to clarify the source of performance gains, and we did not compare with JIT-level optimizations such as Graal or HotSpot intrinsics.

Regarding validity: we tested with arrays of 100  million \ff{int} elements---results may differ with other sizes, types, or objects.
The benchmark by \citet{moller2020eliminating} has low complexity compared to real-world pipelines, lacking method references and autoboxing.
We used a single JVM vendor and hardware configuration with JMH measurements, where JIT behavior variations could affect outcomes.
While Apache Kafka covers nearly one million lines of code, it may not represent all Stream API usage patterns, limiting generalizability to other codebases.

  \section{Discussion}
  \label{sec:discussion}

We found no conclusive evidence of performance gain from optimizing Apache Kafka: total test duration before and after optimization differed by less than 1\% across ten runs lasting over two hours each, with individual test times fluctuating up to five times---consistent with \citet{georges2007statistically} on measurement noise in benchmarking.
In such a mature project (900,000+ lines, 20,000+ pull requests, 1,000+ contributors), developers likely avoid the Stream API in performance-criti\-cal paths, knowing that pipelines are generally slower than imperative loops.

Stream pipelines improve expressiveness and readability, which correlates with higher maintainability and fewer defects---yet these advantages are often sacrificed for performance.
As Apache Cassandra developers noted: ``we quickly realized that Streams and Lambdas were pretty bad from a performance point of view; due to this fact, we stopped using them in hot path''~\cite{mazinanian2017understanding}.
Compile-time optimizations like \hone{} or \streamliner{} may reconcile this trade-off, enabling programmers to adopt streams more widely while retaining acceptable performance.

While bytecode modifications could be done without an IR (e.g., \streamliner{} uses ASM directly), \phic{} enabled us to express transformations declaratively through rewriting rules, eliminating control-flow boilerplate and reducing inconsistent edits.
Its uniform, formally defined structure for all bytecode entities made the pipeline easier to reason about and extend---new optimizations require only new rewriting rules, written in the same surface syntax as the IR itself, without any modification to the rewriting engine or to \jeo{} and \phino{}.
This shift from imperative engineering to a declarative, rule-based system simplified implementation and improved correctness reasoning.
It is important to emphasize that \phic{}~\citep{bugayenko2021eolang} was not designed for this particular experiment: it is a generic formalism for object-oriented languages and is not tied to Java or to the JVM bytecode.
Its expressions describe objects, attributes, and application in a language-agnostic way, and the mapping to Java bytecode performed by \jeo{} is only one possible front-end; rewriting rules operate on \phic{} expressions and are therefore reusable across any source language that can be translated into this IR.

Our focus on \ff{map} and \ff{filter} is intentional: they form a minimal yet practically relevant subset of the Stream API, and they are the two operations for which \ff{mapMulti} provides a natural fusion target with favorable performance characteristics on modern JVMs.
The goal of this study was not to cover the full surface of the Stream API, but to investigate whether fusion via \ff{mapMulti} is a viable alternative to loop unrolling on recent JVMs, where AOT unrolling no longer yields consistent gains.
The rewriting pipeline itself is not restricted to these two operations: because \phic{} expressions uniformly represent all bytecode entities and transformations are expressed as rewriting rules, supporting additional intermediate operations---such as \ff{reduce}, \ff{flatMap}, \ff{peek}, or \ff{takeWhile}---amounts to adding new rules rather than changing the engine.
We view the present work as a proof of concept for this broader program of \phic{}-based stream optimization, and leave a systematic extension to the remaining intermediate operations for future work.

The absolute numbers---matching \streamliner{} on most micro-benchmarks, outperforming it on two, and yielding less than 1\% on Apache Kafka---should be read as evidence that the method is competitive and safe, not as the main contribution.
Our primary contribution is the method itself: a uniform, declarative \phic{}-based IR in which transformations of object-oriented bytecode are expressed and composed as rewriting rules, instead of being hard-coded as ad-hoc pattern matches over raw instructions.
The \ff{map}/\ff{filter}-to-\ff{mapMulti} optimizer is one concrete instantiation, chosen because it is non-trivial to express with traditional bytecode-rewriting frameworks yet small enough to evaluate end-to-end; the same infrastructure can host more aggressive transformations with larger potential payoff.

  \section{Conclusion}
  \label{sec:conclusion}

Stream unrolling is a powerful but intrusive technique; in modern JVMs, stream fusion via \ff{mapMulti} may provide comparable benefits with less intrusion.
Our method differs from existing approaches by introducing a formal calculus for bytecode semantics: we map bytecode to well-defined structures, apply algebraic rewriting rules, and regenerate optimized bytecode.
Compared to \streamliner{}~\citep{moller2020eliminating} on Zulu JVM~23, two benchmarks took twice as long after their optimization while ours caused no degradation, two benchmarks gained equal performance from both methods, and two benchmarks using \ff{count} and \ff{allMatch} could only be optimized by \streamliner{} as these operations are outside our scope.
We validated correctness on 31,799 Apache Kafka unit tests without disruption.

  \section{Acknowledgement}
  \label{sec:acknowledgement}

The work was sponsored by Huawei.
We are grateful to Nikolai Kudasov for pointing our research into the direction of applying \phic{} to solving the problem of stream fusion in Java.

No parts of the paper were written with the help of AI.
Claude Code was used to fix grammar mistakes.

\hbadness=10000
\makeatletter
\global\@ACM@balancefalse
\renewcommand\@BAdblcol{\if@firstcolumn
       \unvbox\@outputbox \penalty\outputpenalty
       \global\oldvsize=\@colht \global\multiply \@colht by 2
       \global\@BAlanceonetrue
       \global\@firstcolumnfalse
  \else \global\@firstcolumntrue
       \if@BAlanceone
       \global\@BAlanceonefalse\@BAlancecol
       \global\@colht=\oldvsize \fi
     \setbox\@outputbox\vbox to \@colht{\hbox to\textwidth
     {\hbox to\columnwidth {\box\@leftcolumn \hss}\hfil
      \vrule width\columnseprule\hfil \hbox to\columnwidth
      {\box\@outputbox \hss}}\vfil}\@combinedblfloats
     \@outputpage \begingroup \@dblfloatplacement
     \@startdblcolumn \@whilesw\if@fcolmade \fi
     {\@outputpage\@startdblcolumn}\endgroup
  \fi}
\makeatother
\balance
\bibliographystyle{ACM-Reference-Format}
\bibliography{main}

\end{document}